\documentclass[aps,prd,twocolumn,groupedaddress,showpacs,nofootinbib]{revtex4}
\usepackage{mathrsfs,bbm,bm} 
\begin{document}
\title{Stability of QED}
\author{M. P. Fry}
\affiliation{School of Mathematics, University of Dublin, Trinity College, Dublin 2, Ireland}
\date{\today}

\begin{abstract}

It is shown for a class of random, time-independent,
square-integrable, three-dimensional magnetic fields that the
one-loop effective fermion action of four-dimensional QED
increases faster than a quadratic in ${\bf B}$ in the strong coupling limit.
The limit is universal. The result relies on the paramagnetism
of charged spin - $1/2$ fermions and the diamagnetism of charged
scalar bosons.

\end{abstract}


\maketitle

\section{Introduction}

Integrating out the fermion fields in four-dimensional QED
continued to the Euclidean metric results in the measure for the
gauge field integration

\renewcommand{\theequation}{1.\arabic{equation}}

\begin{eqnarray}\label{1.1}
d\mu(A)=Z^{-1}e^{-\int d^4x(\frac14F_{\mu\nu}F_{\mu\nu}+{\rm gauge\ fixing})}\times \nonumber \\
{\rm det_{ren}}(1-eS\hskip-0,08cm\not\hskip-0,13cm A)\,\prod\limits_{x,\mu}\ dA_{\mu}(x),
\end{eqnarray}
where ${\rm det}_{ren}$ is the renormalized fermion determinant defined
in Sec. II; $S$ is the free fermion propagator,
and $Z$ is chosen so that $\int\,d\mu(A)=1$. In the limit
$e=0$ the Gaussian measure for the potential $A_\mu$ is chosen to have
mean zero and covariance

\begin{equation}\label{1.2}
\int\,d\mu(A)A_\mu(x)A_\nu(y)=D_{\mu\nu}(x-y),
\end{equation}
where $D_{\mu\nu}$ is the free photon propagator in some fixed gauge. Naively,
integration over the fermion fields produces the ratio of determinants
${\rm det}(\hskip-0,08cm\not\hskip-0,115cm P+e\hskip-0,08cm\not\hskip-0,13cm A+m)/{\rm det}(\hskip-0,08cm\not\hskip-0,105cm P+m)$ which is not well-defined; 
${\rm det}_{ren}$ makes sense of this ratio. It is gauge invariant and depends only on the field strength $F_{\mu\nu}$ and 
invariants formed from it.

We have chosen to introduce this paper with an abrupt intrusion of definitions 
in order to emphasize the central role of ${\rm det}_{ren}$ in QED: 
it is everywhere. It is the origin of all fermion loops in QED. If there are multiple charged fermions then ${\rm det}_{ren}$ is replaced by a product of renormalization determinants, one for each species. For our purpose here it is sufficient to consider one fermion.

The nonperturbative calculation of ${\rm det}_{ren}$ reduces to finding
the eigenvalues of $S\hskip-0,08cm\not\hskip-0,13cm A$,

\begin{equation}\label{1.3}
\int\,d^4yS(x-y)\hskip-0,08cm\not\hskip-0,13cm A(y)\psi_n(y)=
\frac{1}{e_n}\psi_n(x).
\end{equation}
There are at least two complications. Firstly, $S{\not\hskip-0,12cm A}$ is
not a self-adjoint operator, and so many powerful theorems from
analysis do not apply. And secondly, since $A_\mu$ is part of a
functional measure, it is a random field, making the task of
calculating the $e_n$  for all admissible fields impossible. What can be
done is to expand $\ln\,{\rm det}_{ren}$, the one-loop effective action, in 
a power series in $e$. Then the functional integration can be done term-by-term
to obtain textbook QED.

The first nonperturbative calculation of ${\rm det}_{ren}$ was done by
Heisenberg and Euler \cite{1} seventy five years ago for the
special case of constant electric and magnetic fields. Their paper
gave rise to a vast subfield known as quantum field theory under the
influence of external conditions. A comprehensive review of this
body of work relevant to ${\rm det}_{ren}$ is given by Dunne \cite{2}.

An outstanding problem is the strong field behavior of ${\rm det}_{ren}$
that goes beyond constant fields or slowly varying fields
or special fields rapidly varying in one variable [2, 3].{\footnote{We note here progress in scalar ${\rm QED}_4$ since the review \cite{2} in going beyond these fields. Using the multidimensional worldline instanton technique the vacuum pair production rate has been calculated from the one loop effective action of a charged scalar particle in selected two and three-dimensional electric fields \cite{3}. These fields have to be sufficiently regular in order to define a formal functional semiclassical expansion of the quantum mechanical path integral representation of the effective action. The extension of this technique to spinor QED has not been done yet.}} That is,
what is the strong field behavior of ${\rm det}_{ren}$ for a class of random
fields $F_{\mu\nu}$ on ${\mathbbm R}^4$? What if $\ln\,{\rm det}_{ren}$ increases faster than a quadratic in $F_{\mu\nu}$
for such fields? Is ${\rm det}_{ren}$ integrable for any Gaussian measure in this case? This is a question with profound
implications for the stability of QED in isolation. Of course, QED
is part of the standard model, thereby making the overall stability
question a much more intricate one. Nevertheless, the stability of
QED in isolation remains unknown and deserves an answer.

In this paper we consider the case of square integrable,
time-independent magnetic fields ${\bf B}(x)$ defined on ${\mathbbm R}^3$. 
There are additional technical conditions on ${\bf B}$ introduced later.
The magnetic field lines are typically twisted, tangled loops. We find
that

\begin{equation}\label{1.4}
\lim_{e\to\infty}\frac{\ln\,{\rm det}_{ren}}{e^2\ln\,e}=\frac{||{\bf B}||^2T}{24\pi^2},
\end{equation}
where $||{\bf B}||^2=\int d^3x\,{\bf B}\cdot{\bf  B}(x)$, and $T$ is the size of 
the time box. Since $e$ always multiplies ${\bf B}$, this means that 
$\ln\,{\rm det}_{ren}$ is growing faster
than a quadratic in ${\bf B}$. In the constant field case this result
is formally equivalent to the Heisenberg-Euler result \cite{1}
and to calculations relating the effective Lagrangian to the
short-distance behavior of QED via its perturbative $\beta$-function \cite{2}.
What is notable here is that the strong coupling limit of $\ln\,{\rm det}_{ren}$is universal.

To achieve universality the derivation of (\ref{1.4}) must rely on
general principles. One of these is the conjectured "diamagnetic"
inequality for Euclidean three-dimensional QED, namely

\begin{equation}\label{1.5}
\left|{\rm det}_{{\rm QED}_3}(1-eS\hskip-0,08cm\not\hskip-0,13cm A)\right|\le 1.
\end{equation}

The fermion determinant in (\ref{1.5}) is defined in Sec.II. The
diamagnetic inequality is known to be true for lattice formulations
of ${\rm QED}_3$ obeying reflection positivity and using Wilson fermions \cite{4,5,6}. Since Wilson fermions are CP invariant there is no Chern-Simons term
to interfere with the uniqueness of ${\rm det}_{{\rm QED}_3}$ \cite{7}. And since ${\rm det}_{{\rm QED}_3}$ is gauge invariant there are no divergences when the 
lattice spacing for the fermions is sent to zero. As stated by Seiler \cite{6}, (\ref{1.5}) is more an obvious truth than a conjecture.

Since $\left.{\rm det}_{{\rm QED}_3}\right|_{e=0}=1$ and ${\rm det}_{{\rm QED}_3}$ has no zeros in $e$ for real values
of $e$ when $m\ne 0$ \cite{8}, (\ref{1.5}) can be rewritten as

\begin{equation}\label{1.6}
0<{\rm det}_{{\rm QED}_3}\le 1.
\end{equation}
An inspection of Eq.(\ref{2.4}) below indicates that (\ref{1.6}) is a reflection of the tendency of an external magnetic field to lower the energy of
a charged fermion. Therefore, the historic heading of (\ref{1.5}) and
(\ref{1.6}) as "diamagnetic" inequalities is a misnomer; paramagnetic
inequalities would be a more accurate designation. The detailed
justification for going from (\ref{1.5}) to (\ref{1.6}) is given in Sec.II.

The second general principle underlying (\ref{1.4}) is the diamagnetism
of charged spin-0 bosons in an external magnetic field. This is
encapsulated in one of the versions of Kato's inequality discussed
in Sec. III.

The final essential input to (\ref{1.4}) is a restriction on the
class of fields needed to obtain the limit. These restrictions are
summarized in Sec. IV. As the foregoing remarks indicate, ${\rm QED}_3$ is 
central to the derivation of (\ref{1.4}), and it is to the connection between 
${\rm QED}_3$ and ${\rm QED}_4$ that we now turn.

\vskip0,5cm

\section{QED$_3$ AND QED$_4$}

\subsection{The connection}

The connection has been dealt with previously \cite{9}. In order to
make this paper reasonably self-contained we will review the
relevant definitions and results. The upper bound on ${\rm det}_{ren}$ obtained
in \cite{9} is not optimal; it will be optimized here.

The renormalized and regularized fermion determinant in
Wick-rotated Euclidean ${\rm QED}_4$  with on-shell renormalization, 
${\rm det}_{ren}$, may be defined by Schwinger's proper time 
representation \cite{10}

\setcounter{equation}{0}
\renewcommand{\theequation}{2.\arabic{equation}}

\begin{eqnarray}\label{2.1}
\ln\,{\rm det}_{ren}(1-eS\hskip-0,08cm\not\hskip-0,13cm A)=\frac12\int\limits_0^\infty\frac{dt}{t}\left({\rm Tr}\left\{e^{-P^2t}-\right.\right. \nonumber \\
\exp\left[-(D^2+\frac{e}{2}\sigma_{\mu\nu}F_{\mu\nu})t\right]\biggr\}+\left.\frac{e^2||F||^2}{24\pi^2}\right)e^{-tm^2},
\end{eqnarray}
where $D_\mu=P_\mu-eA_\mu,\ \sigma_{\mu\nu}=(1/2i)[\gamma_\mu,\gamma_\nu],\ \gamma_\mu^\dagger=-\gamma_\mu,\ ||F||^2=\displaystyle\int d^4 xF_{\mu\nu}^2(x)$,
and $e$ is assumed to be real. We choose the chiral representation
of the $\gamma$-matrices so that $\sigma_{ij}=\left(\begin{array}{cc}
-\sigma_k & 0 \\
      0 & -\sigma_k
\end{array}\right)$, $i,j,k=1,2,3$ in cyclic
order. Since we will consider time-independent magnetic fields we
set $A_\mu=(0,{\bf A}(x))$ with $x$ in ${\mathbbm R}^3$. Then (\ref{2.1}) reduces to

\begin{eqnarray}\label{2.2}
\ln\,{\rm det}_{ren}=\frac T2\int\limits_0^\infty\frac{dt}{t}\left[\frac{2}{(4\pi t)^{1/2}}{\rm Tr}\left(e^{-P^2t}-\right.\right. \nonumber \\
\left.\exp\left\{-[({\bf P}-e{\bf A})^2-e{\bm\sigma}\cdot{\bf B}]t\right\}\biggr)\hskip-0,03cm+\frac{e^2||{\bf B}||^2}{12\pi^2}\right]\hskip-0,1cm e^{-tm^2},\hskip0,3cm
\end{eqnarray}
where $T$ is the dimension of the time box, and the factor 2 is
from the partial spin trace. Clearly we must have ${\bf B}\in L^2({\mathbbm R}^3)$. If ${\bf A}$ is assumed to be in the Coulomb gauge ${\bm\nabla}\cdot{\bf A}=0$, then by the Sobolev-Talenti-Aubin inequality \cite{11}

\begin{equation}\label{2.3}
\int d^3x{\bf B}(x)\cdot{\bf B}(x)\ge\left(\frac{27\pi^4}{16}\right)^{1/3}\sum_{i=1}^3\left(\int d^3x|A_i(x)|^6\right)^{1/3}.
\end{equation}
So we must also have ${\bf A}\in L^6({\rm I\hskip-0,06cm R^3})$.

In analogy with ${\rm det}_{ren}$ in (\ref{2.1}), without the charge
renormalization subtraction, ${\rm det}_{{\rm QED}_3}$ may be defined by

\begin{eqnarray}\label{2.4}
\ln\,{\rm det}_{{\rm QED}_3}(m^2)=\frac12\int\limits_0^\infty\frac{dt}{t}{\rm Tr}\left(e^{-P^2t}-\right. \nonumber \\
\exp\left\{-[({\bf P}-e{\bf A})^2-e{\bm\sigma}\cdot{\bf B}]t\right\}\biggr)e^{-tm^2}.
\end{eqnarray}
This definition and regularization of ${\rm det}_{{\rm QED}_3}$ is parity 
conserving and gives no Chern-Simons term. Substituting (\ref{2.4}) in 
(\ref{2.2}) and,
noting that $\pi^{-1}\int\limits_0^\infty dE\,e^{-tE^2}=(4\pi t)^{-1/2}$,        we obtain \cite{9}

\begin{equation}\label{2.5}
\begin{array}{ccc}
\ln\,{\rm det}_{ren}=\displaystyle\frac{2T}{\pi}\int\limits_0^\infty dE\biggl(\ln\,{\rm det}_{{\rm QED}_3}(E^2+m^2) \\
+\displaystyle\frac{e^2||{\bf B}||^2}{24\pi^{3/2}}\int\limits_0^\infty\frac{dt}{t^{1/2}}e^{-(E^2+m^2)t}\biggr) \\
=\displaystyle\frac{T}{\pi}\int\limits_{m^2}^\infty\frac{dM^2}{\sqrt{M^2-m^2}}\left(\ln\,{\rm det}_{{\rm QED}_3}(M^2)+\frac{e^2||{\bf B}||^2}{24\pi\sqrt{M^2}}\right).
\end{array}
\end{equation}
Result (\ref{2.5}) will be referred to repeatedly in what follows.

\vskip0,5cm

\subsection{Justification of (\ref{1.6})}

Continuing our review of previous work we turn to the
derivation of the upper bound on $\ln\,{\rm det}_{ren}$ in (\ref{1.4}). 
Since the degrees of divergence of the first, second and third-order
contributions to $\ln\,{\rm det}_{{\rm QED}_3}$ are 2,1 and 0, respectively, 
these must be dealt with separately. Their definition is obtained
from the expansion of (\ref{2.4}) through $O(e^3)$, resulting in

\begin{eqnarray}\label{2.6}
\ln\,{\rm det}_{{\rm QED}_3}(1-eS\hskip-0,09cm\not\hskip-0,13cm A)=-\frac{e^2}{4\pi}\int\frac{d^3k}{(2\pi)^3}|{\hat{\bf B}}(k)|^2 \nonumber \\
\times\int\limits_0^1dz\frac{z(1-z)}{[z(1-z)k^2+m^2]^{1/2}}+\ln\,{\rm det}_4(1-eS\hskip-0,09cm\not\hskip-0,13cm A),
\end{eqnarray}
where $\ln\,{\rm det}_4$  defines the remainder and $\hat{\bf B}$ is the Fourier transform of ${\bf B}$. Definition (\ref{2.4}) assigns the
value of zero to the terms of order $e$ and $e^3$. The argument of 
${\rm det}_{{\rm QED}_3}$ has been changed to indicate its origin as the formal ratio of ${\rm QED}_3$ determinants ${\rm det}(\not\hskip-0,11cm P-e\not\hskip-0,16cm A+m)/$ ${\rm det}(\not\hskip-0,11cm P+m)$. Note the minus sign
in (\ref{2.6}) pointing to paramagnetism.

The following theorems are essential for what follows:

Theorem 1 \cite{5,12,13}. Let the operator $S{\not\hskip-0,12cm A}$ in ${\rm det}_4$ be transformed by a similarity transformation to $K=(p^2+m^2)^{1/4}S{\not\hskip-0,12cm A}(p^2+m^2)^{-1/4}$. This leaves the eigenvalues of $S{\not\hskip-0,12cm A}$ invariant. Then $K$ is a bounded operator on $L^2({\mathbbm R}^3,d^3x;{\mathbbm C}^2)$ for ${\bf A}\in L^p({\mathbbm R}^3)$ for $p>3$. Moreover, $K$ is a compact operator belonging to the trace ideal ${\mathscr I}_p,\ p>3$.

The trace ideal ${\mathscr I}_p(1\le p<\infty)$ is defined as those compact
operators $A$ with  $||A||_p^p={\rm Tr}((A^\dag A)^{p/2})<\infty$. From this it follows that the eigenvalues $1/e_n$ of $S\hskip-0,09cm\not\hskip-0,13cm A$ obtained from (\ref{1.3}) specialized to three dimensions are of finite multiplicity and satisfy
$\sum\limits_{n=1}^{\infty}|e_n|^{-p}<\infty$
for $p>3$. The eigenfunctions $\psi_n$ belong to the Sobolev space $L^2({\mathbbm R}^3,\sqrt{k^2+m^2}\,d^3k:{\mathbbm C})$. None of the $e_n$ are real for $m\ne 0$ \cite{8}.

Theorem 2 \cite{14,15,16}. Define the regularized determinant

\begin{equation}\label{2.7}
{\rm det}_n(1+A)={\rm det}\left[(1+A)\exp\left(\sum_{k=1}^{n-1}(-1)^kA^k/k\right)\right].
\end{equation}
Then ${\rm det}_n$ can be expressed in terms of the eigenvalues of $A\in{\mathscr I}_p$ for $n\ge p$.

Accordingly, ${\rm det}_4$ in (\ref{2.6}) is defined and can be represented
as \cite{16}

\begin{equation}\label{2.8}
{\rm det}_4(1-eS\hskip-0,09cm\not\hskip-0,13cm A)=\prod\limits_{n=1}^{\infty}\left[\left(1-\frac{e}{e_n}\right)\exp\left(\sum_{k=1}^3\left(\frac{e}{e_n}\right)^k/k\right)\right].
\end{equation}
The reality of ${\rm det}_4$ for real $e$ and $C$-invariance require that the
eigenvalues $e_n$ appear in the complex plane as quartets $\pm e_n,\ \pm e_n^*$ or as imaginary pairs when $m\ne 0$. As expected, the expansion of $\ln\,{\rm det}_4$ in powers of $e$ begins in fourth order. 

We have established that ${\rm det}_4|_{e=0}=1$ and that ${\rm det}_4$ has no zeros for real values of $e$. Therefore, by (\ref{2.6}) ${\rm det}_{{\rm QED}_3}>0$ for all real $e$, thereby allowing one to go from ({\ref{1.5}) to (\ref{1.6}). It might be objected that this is obvious, but we will need the detailed information introduced about ${\rm det}_4$ in the sequel.

The determinant ${\rm det}_4$ is an entire function of $e$ considered as
a complex variable, meaning that it is holomorphic in the entire
complex $e$-plane. Since $\sum\limits_{n=1}^{\infty}|e_n|^{-3-\epsilon}<\infty$ for $\epsilon>0$, its order is at most 3 \cite{15,17}. 
This means that for any complex value of $e$, and
positive constants $A,K,\ |{\rm det}_4|<A(\epsilon)\exp(K(\epsilon)|e|^{3+\epsilon})$ for any $\epsilon>0$.

From (\ref{1.6}), (\ref{2.6}) and for real values of $e$
\begin{equation}\label{2.9}
\ln\,{\rm det}_4\le\frac{e^2}{4\pi}\int\frac{d^3k}{(2\pi)^3}|\hat{\bf B}(k)|^2\int\limits_0^1dz\frac{z(1-z)}{[z(1-z)k^2+m^2]^{1/2}}.
\end{equation}
This is a truly remarkable inequality. Referring to (\ref{2.9}), 
${\rm det}_4$'s growth is slower on the real $e$-axis than its potential growth
in other directions. We also note that ${\rm det}_4$ is largely unknown. Even
the reduction of the fourth-order term in its expansion to an
explicitly gauge invariant form involving only ${\bf B}$-fields requires a
huge effort when the fields are not constant \cite{18}. The sixth-order
reduction has not been completed as far as the author knows.

\vskip0,5cm

\subsection{Upper bound on ${\rm\bf det}_{ren}$}

Insert (\ref{2.6}) in (\ref{2.5}) and get

\begin{eqnarray}\label{2.10}
\ln\,{\rm det}_{ren}=\frac{e^2T}{4\pi^2}\int\frac{d^3k}{(2\pi)^3}|{\hat{\bf B}}(k)|^2\int\limits_0^\infty dz\,z(1-z) \nonumber \\
\times\ln\left[\frac{z(1-z)k^2+m^2}{m^2}\right]\hskip-0,1cm+\frac{T}{\pi}\int\limits_{m^2}^{\infty}\hskip-0,1cm\frac{dM^2}{\sqrt{M^2-m^2}}\ln{\rm det}_4(M^2).\nonumber \\
\quad
\end{eqnarray}
The objective here is to obtain the behavior of $\ln\,{\rm det}_{ren}$ when
the coupling $e$ is large, real and positive. Since $e$ always multiplies ${\bf B}$ we introduce the scale parameter ${\mathscr B}=\max\limits_x|{\bf B}|$, which has the dimension of $M^2$. Why ${\mathscr B}$ is finite will be explained in Sec. III.B. Then the integral in (\ref{2.10}) is broken up into $\int\limits_{m^2}^{e{\mathscr B}}$ and $\int\limits_{e{\mathscr B}}^{\infty}$. 

Substitution of (\ref{2.9}) into the lower range integral gives

\begin{eqnarray}\label{2.11}
\ln\,{\rm det}_{ren}\le\frac{e^2T}{4\pi^2}\int\frac{d^3k}{(2\pi)^3}|\hat{\bf B}(k)|^2\int\limits_0^1dz\,z(1-z) \nonumber \\
\times\ln\left(\frac{4e{\mathscr B}+2z(1-z)k^2-2m^2}{m^2}\right) \nonumber \\
+\frac{T}{\pi}\int\limits_{e{\mathscr B}}^{\infty}\frac{dM^2}{\sqrt{M^2-m^2}}\ln\,{\rm det}_4(e{\bf B},M^2).
\end{eqnarray}
We have simplified the argument of the logarithm using
$2\sqrt{xy}\le x+y$ for $x,y\ge 0$. Then for $e{\mathscr B}\gg m^2$

\begin{eqnarray}\label{2.12}
\ln\,{\rm det}_{ren}\le\frac{e^2T||{\bf B}||^2}{24\pi^2}\ln\left(\frac{4e{\mathscr B}}{m^2}\right) \nonumber \\
+\frac{T}{\pi}\int\limits_{e{\mathscr B}}^{\infty}\frac{dM^2}{\sqrt{M^2-m^2}}\ln\,{\rm det}_4(e{\bf B},M^2) \nonumber \\
+O\left(\frac{eT\int d^3x{\bf B}\cdot\nabla^2{\bf B}}{{\mathscr B}}\right).
\end{eqnarray}
The integral in (\ref{2.12}) can be estimated by making a large mass
expansion of $\ln\,{\rm det}_4$. This is facilitated by inserting (\ref{2.6}) in
(\ref{2.4}) and examining the small $t$ region of $\ln\,{\rm det}_4$'s resulting proper time representation. The details of this expansion are
in Sec. 3B of \cite{9}, and give the result

\begin{eqnarray}\label{2.13}
\ln\,{\rm det}_4(e{\bf B},M^2){{{\phantom{\sqrt{\hat{OSR}}}}\atop{{=}\atop{\scriptscriptstyle M\to\infty}}}}\frac12\int\limits_0^\infty\frac{dt}{t}(4\pi t)^{-3/2}e^{-tM^2} \nonumber \\
\times\int d^3x\left[\frac{2}{45}e^4t^4({\bf B}\cdot{\bf B})^2+O(e^4t^5{\bf B}\cdot{\bf B}{\bf B}\cdot\nabla^2{\bf B})\right] \nonumber \\
=\frac{e^4\int({\bf B}\cdot{\bf B})^2}{480\pi M^5}+O\left(\frac{e^4\int{\bf B}\cdot{\bf B}{\bf B}\cdot\nabla^2{\bf B}}{M^7}\right).\quad
\end{eqnarray}

In the first line of (\ref{2.13}) it is assumed that the heat kernel
expansion is an asymptotic expansion in $t$ in the strict sense of
its definition, namely \cite{19}

\begin{eqnarray}\label{2.14}
<x|e^{-t[({\bf P}-e{\bf A})^2-e{\bm\sigma}\cdot{\bf B}]}|x>-(4\pi t)^{-3/2}\sum_{n=0}^Na_n(x)t^n \nonumber \\
{{\phantom{qwqwy}}\atop{\widetilde{\scriptscriptstyle t\to 0^+}}}(4\pi t)^{-3/2}a_{N+1}(x)t^{N+1}.\qquad\quad
\end{eqnarray}
This must hold for every $N$. A necessary condition for (\ref{2.14}) is
that ${\bf B}$ be infinitely differentiable to ensure that each coefficient
$a_n$ is finite. As far as the author knows it is not known yet if
this is a sufficient condition. So (\ref{2.14}) is an assumption that may
require additional conditions on ${\bf B}$. Only coefficients $a_n$ of $O(e^{2n}),\ n\ge 2$ are present in $\ln\,{\rm det}_4$'s expansion.

The t-integration in (\ref{2.13}), although extending to infinity,
is limited to small $t$ since $M\to\infty$ due to the parameter $e{\mathscr B}$ in
(\ref{2.12}). Substituting (\ref{2.13}) in (\ref{2.12}) results in

\begin{eqnarray}\label{2.15}
\ln{\rm det}_{ren}\le\frac{e^2||{\bf B}||^2T}{24\pi^2}\ln\left(\frac{4e{\mathscr B}}{m^2}\right)+O\left(\frac{e^2T\int({\bf B}\cdot{\bf B})^2}{{\mathscr B}^2}\right) \nonumber \\
+O\left(\frac{eT\int{\bf B}\cdot\nabla^2{\bf B}}{{\mathscr B}}\right),\qquad\qquad\qquad\quad
\end{eqnarray}
or
\begin{equation}\label{2.16}
\lim_{e\to\infty}\frac{\ln\,{\rm det}_{ren}}{e^2\ln\,e}\le\frac{||{\bf B}||^2T}{24\pi^2},
\end{equation}
consistent with (\ref{1.4}). This bound is independent of the
charge renormalization subtraction point. If the subtraction
were made at photon momentum $k^2=\mu^2$ instead of $k^2=0$ then the $\ln\,m^2$
terms in (\ref{2.10}) and (\ref{2.11}) would be replaced with $\ln[z(1-z)\mu^2+m^2]$, which has nothing to do with strong coupling.

The scaling procedure used here is designed to obtain the
least upper bound on $\ln\,{\rm det}_{ren}$.
In \cite{9} we chose to break up the $M$-integral
as $\int\limits_{m^2}^{e^4||{\bf B}||^4}$ and $\int\limits_{e^4||{\bf B}||^4}^{\infty}$. This resulted in a fast $1/e^4$ falloff
of the $\ln\,{\rm det}_4$ terms compared to $e^2$ here, but gave a
weaker upper bound on $\ln\,{\rm det}_{ren}$, namely

\begin{equation}\label{2.17}
\lim_{e\to\infty}\frac{\ln\,{\rm det}_{ren}}{e^2\ln\,e}\le\frac{||{\bf B}||^2T}{6\pi^2}.
\end{equation}
We mention that the coefficient 1/960 in (\ref{3.17}) in \cite{9} should
be 1/360.

Here we might have chosen a more general scaling such as $e^{\alpha}(\ln e)^{\beta}{\mathscr B}$ or $e^{\alpha}(\ln\ln e)^{\beta}{\mathscr B}$, etc., with $\alpha\ge 1,\,\beta>0$. Then the right-hand side of (\ref{2.16}) would have been replaced with $\alpha||{\bf B}||^2T/24\pi^2$. The case $\alpha<1$ causes the first remainder term in (\ref{2.15}) to be no longer subdominant. Therefore, our scaling $e{\mathscr B}$ is an optimal one.

\vskip0,5cm

\section{Lower bound on ${\rm\bf det}_{ren}$}

\subsection{Fundamentals}

On referring to (\ref{2.5}) the lower bound on ${\rm det}_{ren}$ will come
from operations on $\ln{\rm det}_{{\rm QED}_3}$. We begin with the operator
identity (A2) in Appendix A applied to $\ln{\rm det}_{{\rm QED}_3}$ in (\ref{2.4}). Letting $X=({\bf P}-e{\bf A})^2$ and $Y=-e{\bm\sigma}\cdot{\bf B}$ we obtain

\setcounter{equation}{0}
\renewcommand{\theequation}{3.\arabic{equation}}

\begin{equation}\label{3.1}
\begin{array}{cccc}
\ln{\rm det}_{{\rm QED}_3}=\displaystyle-\frac12\int\limits_0^\infty\frac{dt}{t}{\rm Tr}\left(e\int\limits_0^tds\,e^{-(t-s)({\bf P}-e{\bf A})^2}\right.  \\
\displaystyle\times{\bm\sigma}\cdot{\bf B}e^{-s({\bf P}-e{\bf A})^2}\hskip-0,1cm+e^2\hskip-0,1cm\int\limits_0^t\hskip-0,1cm ds_1\hskip-0,2cm\int\limits_0^{t-s_1}\hskip-0,2cm ds_2e^{-(t-s_1-s_2)[({\bf P}-e{\bf A})^2-e{\bm\sigma}\cdot{\bf B}]} \\
\displaystyle\times{\bm\sigma}\cdot{\bf B}e^{-s_2({\bf P}-e{\bf A})^2}{\bm\sigma}\cdot{\bf B}e^{-s_1({\bf P}-e{\bf A})^2}\Biggr)e^{-m^2t} \\
\displaystyle+\frac12\int\limits_0^\infty\frac{dt}{t}{\rm Tr}\left(e^{-{\bf P}^2t}-e^{-({\bf P}-e{\bf A})^2t}\right)e^{-tm^2}.
\end{array}
\end{equation}
The spin trace in the first term is zero, and the last term,
after tracing over spin, is the one-loop effective action of
scalar ${\rm QED}_3$,
\begin{equation}\label{3.2}
\ln{\rm det}_{S{\rm QED}_3}=\int\limits_0^{\infty}\frac{dt}{t}e^{-tm^2}{\rm Tr}\left(e^{-{\bf P}^2t}-e^{-({\bf P}-e{\bf A})^2t}\right).
\end{equation}
Thus,
\begin{eqnarray}\label{3.3}
\ln{\rm det}_{{\rm QED}_3}=\ln{\rm det}_{S{\rm QED}_3}-\frac{e^2}{2}\int\limits_0^\infty\frac{dt}{t}e^{-tm^2}{\rm Tr}\left(\int\limits_0^tds_1\right. \nonumber \\
\times\int\limits_0^{t-s_1}ds_2\ e^{-(t-s_1-s_2)[({\bf P}-e{\bf A})^2-e{\bm\sigma}\cdot{\bf B}]} \qquad\quad \nonumber \\
\times{\bm\sigma}\cdot{\bf B}e^{-s_2({\bf P}-e{\bf A})^2}{\bm\sigma}\cdot{\bf B}e^{-s_1({\bf P}-e{\bf A})^2}\biggr),\qquad
\end{eqnarray}
remembering that the factor $1/2$ in the last term of (\ref{3.1}) is
cancelled by the spin trace.

Let $\Delta_A=[({\bf P}-e{\bf A})^2+m^2]^{-1}$. In Appendix B it is shown that
$\Delta_A^{1/2}{\bm\sigma}\cdot{\bf B}\Delta_A^{1/2}\in{\mathscr I}_2$;
that is, it is a Hilbert-Schmidt operator provided ${\bf B}\in L^2$ and $m\ne 0$. 
Then (\ref{2.7}) gives

\begin{equation}\label{3.4}
\begin{array}{llllll}
\displaystyle\ln{\rm det}_2(1-e\Delta_A^{1/2}{\bm\sigma}\cdot{\bf B}\Delta_A^{1/2}) \\
\displaystyle=\ln{\rm det}\left[(1-e\Delta_A^{1/2}{\bm\sigma}\cdot{\bf B}\Delta_A^{1/2})e^{e\Delta_A^{1/2}{\bm\sigma}\cdot{\bf B}\Delta_A^{1/2}}\right] \\
\displaystyle={\rm Tr}\ln\left[(1-e\Delta_A^{1/2}{\bm\sigma}\cdot{\bf B}\Delta_A^{1/2})e^{e\Delta_A^{1/2}{\bm\sigma}\cdot{\bf B}\Delta_A^{1/2}}\right] \\
\displaystyle=\hskip-0,05cm{\rm Tr}\hskip-0,1cm\left[\hskip-0,02cm\int\limits_0^{\infty}\hskip-0,1cm\frac{dt}{t}e^{-tm^2}\hskip-0,15cm\left(\hskip-0,1cm e^{-({\bf P}-e{\bf A})^2t}\hskip-0,15cm-\hskip-0,1cm e^{-[({\bf P}-e{\bf A})^2-e{\bm\sigma}\cdot{\bf B}]t}\hskip-0,05cm\right)\hskip-0,1cm+\hskip-0,05cm e\Delta_A{\bm\sigma}\hskip-0,1cm\cdot\hskip-0,1cm{\bf B}\hskip-0,05cm\right] \\
\displaystyle=-e^2\hskip-0,1cm\int\limits_0^{\infty}\hskip-0,1cm\frac{dt}{t}e^{-tm^2}\hskip-0,2cm\int\limits_0^t\hskip-0,1cm ds_1\hskip-0,2cm\int\limits_0^{t-s_1}\hskip-0,2cm ds_2{\rm Tr}\left(e^{-(t-s_1-s_2)[({\bf P}-e{\bf A})^2-e{\bm\sigma}\cdot{\bf B}]}\right. \\
\displaystyle\times{\bm\sigma}\cdot{\bf B}e^{-s_2({\bf P}-e{\bf A})^2}{\bm\sigma}\cdot{\bf B}e^{-s_1({\bf P}-e{\bf A})^2}\biggr).\quad
\end{array}
\end{equation}
In going from the penultimate to the last line in (\ref{3.4}) use was
again made of the identity (A2). Substituting (\ref{3.4}) in (\ref{3.3})
gives
\begin{equation}\label{3.5}
\ln{\rm det}_{{\rm QED}_3}=\frac12\ln{\rm det}_2(1-e\Delta_A^{1/2}{\bm\sigma}\cdot{\bf B}\Delta_A^{1/2})+\ln{\rm det}_{S{\rm QED}_3}.
\end{equation}
As $\ln{\rm det}_{{\rm QED}_3}$ and $\ln{\rm det}_2$ are well-defined by our choice of fields, so is $\ln{\rm det}_{{\rm SQED}_3}$ in ({\ref{3.5}). What has been accomplished here
is to isolate the Zeeman term ${\bm\sigma}\cdot{\bf B}$ in 
$\ln{\rm det}_2$. Since $\Delta_A^{1/2}{\bm\sigma}\cdot{\bf B}\Delta_A^{1/2}$ is Hilbert-Schmidt and self-adjoint, 
$\ln{\rm det}_2$ is susceptible to extensive analytic analysis.

Substitute (\ref{3.5}) in (\ref{2.5}):
\begin{eqnarray}\label{3.6}
\ln{\rm det}_{ren}=\frac{T}{\pi}\int\limits_{m^2}^{\infty}\frac{dM^2}{\sqrt{M^2-m^2}} \nonumber \\
\times\left(\frac12\ln{\rm det}_2(1-e\Delta_A^{1/2}{\bm\sigma}\cdot{\bf B}\Delta_A^{1/2})\right. \nonumber \\
+\ln{\rm det}_{S{\rm QED}_3}+\frac{e^2||{\bf B}||^2}{24\pi\sqrt{M^2}}\biggr).
\end{eqnarray}
We now introduce two central inequalities. The first relies on
the diamagnetism of charged scalar bosons as expressed by
Kato's inequality in the form \cite{20,21}
\begin{equation}\label{3.7}
{\rm Tr}\left(e^{-({\bf P}-e{\bf A})^2t}\right)\le{\rm Tr}\ e^{-{\bf P}^2t}.
\end{equation}
This implies that on average the energy eigenvalues of such
bosons rise in a magnetic field and hence by(\ref{3.2}) that \cite{21}
\begin{equation}\label{3.8}
\ln{\rm det}_{S{\rm QED}_3}\ge 0.
\end{equation}

The second inequality is introduced beginning with the
penultimate line of (\ref{3.4}). Noting that the spin trace of the
$\Delta_A{\bm\sigma}\cdot{\bf B}$ term is zero, then
\begin{eqnarray}\label{3.9}
\ln{\rm det}_2\left(1-e\Delta_A^{1/2}{\bm\sigma}\cdot{\bf B}\Delta_A^{1/2}\right)=\int\limits_0^\infty\frac{dt}{t}e^{-tm^2} \nonumber \\
\times{\rm Tr}\left(e^{-t({\bf P}-e{\bf A})^2}-e^{-[({\bf P}-e{\bf A})^2-e{\bm\sigma}\cdot{\bf B}]t}\right).
\end{eqnarray}
By the Bogoliubov-Peierls inequality \cite{22,23} and Sec. 2.1,8 of \cite{24}
\begin{equation}\label{3.10}
{\rm Tr}\,e^{-[({\bf P}-e{\bf A})^2-e{\bm\sigma}\cdot{\bf B}]t}\ge{\rm Tr}\,e^{-t({\bf P}-e{\bf A})^2t}\,e^{-te<{\bm\sigma}\cdot{\bf B}>},
\end{equation}
where
\begin{equation}\label{3.11}
<{\bm\sigma}\cdot{\bf B}>=\frac{{\rm Tr}\left({\bm\sigma}\cdot{\bf B}e^{-t({\bf P}-e{\bf A})^2t}\right)}{{\rm Tr}\,e^{-({\bf P}-e{\bf A})^2t}}=0.
\end{equation}
Hence,
\begin{equation}\label{3.12}
\ln{\rm det}_2\left(1-e\Delta_A^{1/2}{\bm\sigma}\cdot{\bf B}\Delta_A^{1/2}\right)\le 0.
\end{equation}
consistent with (\ref{3.5}) when combined with (\ref{1.6}) and (\ref{3.8}).

There is another reason why (\ref{3.12}) holds. Let $C=e\Delta_A^{1/2}{\bm\sigma}\cdot{\bf B}\Delta_A^{1/2}$. Since $C$ is Hilbert-Schmidt,
\begin{eqnarray}\label{3.13}
\ln{\rm det}_2(1-C)&=&\ln{\rm det}\left[(1-C)e^C\right] \nonumber \\
                   &=&{\rm Tr}\left[\ln(1-C)+C\right] \nonumber \\
                   &=&\frac12{\rm Tr}\ln(1-C^2) \nonumber \\
                   &=&\frac12\sum\limits_{n=1}^{\infty}\ln(1-\lambda_n^2).
\end{eqnarray}
The third line of (\ref{3.13}) follows from the second since
the trace over spin eliminates all odd powers of $C$. In the
last line we introduced the real eigenvalues $\lambda_n$ of
$e\Delta_A^{1/2}{\bm\sigma}\cdot{\bf B}\Delta_A^{1/2}$. Since 
$\ln{\rm det}_2$ is real and finite then
$|\lambda_n|<1$ for all $n$, giving (\ref{3.12}). Because
$\Delta_A^{1/2}{\bm\sigma}\cdot{\bf B}\Delta_A^{1/2}\in{\mathscr I}_2$, 
it is a compact operator, and so the $\lambda_n$ are countable and of
finite multiplicity. 

Now consider
\begin{eqnarray}\label{3.15}
\frac{\partial}{\partial m^2}\ln{\rm det}_2(m^2)=\int\limits_0^{\infty}dt\,e^{-tm^2} \nonumber \\
\times{\rm Tr}\left(e^{-[({\bf P}-e{\bf A})^2-e{\bm\sigma}\cdot{\bf B}]t}-e^{-({\bf P}-e{\bf A})^2t}\right)\ge 0,
\end{eqnarray}
by (\ref{3.9})-(\ref{3.11}). Therefore, ${\rm det}_2$ is a monotonically increasing function of $m^2$  .

Next, break up the $M$-integral in (\ref{3.6}) as in Sec.II.C:
\begin{equation}\label{3.16}
\begin{array}{ccc}
\displaystyle\ln{\rm det}_{ren}\hskip-0,05cm=\hskip-0,05cm\frac{T}{\pi}\hskip-0,1cm\int\limits_{m^2}^{e{\mathscr B}}\hskip-0,2cm\frac{dM^2}{\sqrt{M^2-m^2}}\hskip-0,1cm\left(\frac12\ln{\rm det}_2(1-e\Delta_A^{1/2}{\bm\sigma}\hskip-0,05cm\cdot\hskip-0,05cm{\bf B}\Delta_A^{1/2})\right. \\
\displaystyle+\ln{\rm det}_{S{\rm QED}_3}+\frac{e^2||{\bf B}||^2}{24\pi\sqrt{M^2}}\biggr)\hskip2cm  \\
\displaystyle\hskip1,2cm+\frac{T}{\pi}\int\limits_{e{\mathscr B}}^{\infty}\frac{dM^2}{\sqrt{M^2-m^2}}\left(\ln{\rm det}_{{\rm QED}_3}+\frac{e^2||{\bf B}||^2}{24\pi\sqrt{M^2}}\right),
\end{array}
\end{equation}
where we reinserted (\ref{3.5}) into the upper-range $M$-integral. 
By (\ref{3.15})
\begin{equation}\label{3.17}
\begin{array}{ll}
\displaystyle\hskip-0,1cm\int\limits_{m^2}^{e{\mathscr B}}\hskip-0,2cm\frac{dM^2}{\sqrt{M^2-m^2}}\ln{\rm det}_2(M^2)\ge\ln{\rm det}_2\biggr|_{\scriptscriptstyle M^2=m^2}\hskip-0,1cm\int\limits_{m^2}^{e{\mathscr B}}\hskip-0,2cm\frac{dM^2}{\sqrt{M^2-m^2}} \\
\displaystyle=2\ln{\rm det}_2\left(1-e\Delta_A^{1/2}{\bm\sigma}\cdot{\bf B}\Delta_A^{1/2}\right)\biggr|_{\scriptscriptstyle M^2=m^2}\sqrt{e{\mathscr B}-m^2}. 
\end{array}
\end{equation}
Hence, (\ref{3.8}) and (\ref{3.17}) result in (\ref{3.16}) becoming
\begin{equation}\label{3.18}
\begin{array}{lll}
\displaystyle\ln{\rm det}_{ren}\ge\frac{T}{\pi}\sqrt{e{\mathscr B}-m^2}\ln{\rm det}_2\hskip-0,1cm\left(\hskip-0,1cm 1\hskip-0,1cm-e\Delta_A^{1/2}{\bm\sigma}\hskip-0,05cm\cdot\hskip-0,05cm{\bf B}\Delta_A^{1/2}\right)\hskip-0,1cm\biggr|_{\scriptscriptstyle M^2=m^2}  \\
\displaystyle+\frac{e^2T||{\bf B}||^2}{24\pi^2}\ln\left(\frac{e{\mathscr B}}{m^2}\right)+\frac{e^2T}{12\pi^2}||{\bf B}||^2\ln\left(1+\sqrt{1-\frac{m^2}{e{\mathscr B}}}\right) \\
\displaystyle+\frac{T}{\pi}\int\limits_{e{\mathscr B}}^{\infty}\frac{dM^2}{\sqrt{M^2-m^2}}\left(\ln{\rm det}_{{\rm QED}_3}+\frac{e^2||{\bf B}||^2}{24\pi\sqrt{M^2}}\right). 
\end{array}
\end{equation}
We now turn to the strong coupling behavior of $\ln{\rm det}_2$.

\vskip0,5cm

\subsection{Strong coupling behavior of ${\rm\bf ln\ det_2}$}

The eigenvalues $\lambda_n$ in (\ref{3.13}) are obtained from
\begin{equation}\label{3.19}
e\Delta_A^{1/2}{\bm\sigma}\cdot{\bf B}\Delta_A^{1/2}\varphi_n=\lambda_n\varphi_n,
\end{equation}
for $\varphi_n\in L^2$ following the remark under (\ref{B3}) in Appendix B. Letting
$\Delta_A^{1/2}\varphi_n=\psi_n$ gives
\begin{equation}\label{3.20}
\left[({\bf P}-e{\bf A})^2-\frac{e{\bm\sigma}\cdot{\bf B}}{\lambda_n}\right]\psi_n=-m^2\psi_n,
\end{equation}
where $\psi_n\in L^2$ provided $m\ne 0$. This follows from (B5) and Young's
inequality (B7). The requirement that $m\ne 0$ follows from the role
of the eigenvalues $\{\lambda_n\}_{n=1}^{\infty}$ as adjustable coupling constants whose discrete values result in bound states with energy $-m^2$  for a fixed
value of $e$. Since the operator $({\bf P}-e{\bf A})^2-e{\bm\sigma}\cdot{\bf B}\ge 0$, such bound states are impossible unless 
$|\lambda_n|<1$ for all $n$, which is the
physical reason why (\ref{3.12}) is true. Inspection of (\ref{3.20}) suggests
that as $e$ increases $|\lambda _n|$ likewise increases for fixed $n$ to maintain the bound state energy at $-m^2$  . This is illustrated by the constant
field case that is excluded from our analysis:
\begin{equation}\label{3.21}
|\lambda_n|=\frac{|eB|}{(2n+1)|eB|+m^2},\ n=0,1,\ldots.
\end{equation}

Because the operator $\Delta_A^{1/2}{\bm\sigma}\cdot{\bf B}\Delta_A^{1/2}$      is Hilbert-Schmidt the eigenfunction $\varphi_n$ has finite multiplicity, and 
the $\lambda_n$ in (\ref{3.13}) are counted up to this multiplicity. To estimate the multiplicity note that the eigenfunctions $\varphi_n$ and $\psi_n$ are in
one-to-one correspondence. Next, note that for $\psi\in L^2 ({\mathbbm R}^3;{\mathbbm C}^2)$ and a generic $\lambda$ with $|\lambda|<1$,
\begin{equation}\label{3.22}
(\psi,[({\bf P}-e{\bf A})^2-\frac{e}{\lambda}{\bm\sigma}\cdot{\bf B}]\psi)\ge(\psi,[({\bf P}-e{\bf A})^2-\left|\frac{e}{\lambda}\right||{\bf B}|)\psi.
\end{equation}
Thus the Hamiltonian on the left, $H_+$, dominates that on the
right, $H_-$. Let $N_{-m^2}(H)$ denote the dimension of the spectral
projection onto the eigenstates of Hamiltonian $H$ with eigenvalues
less than or equal to $-m^2$. Because $H_+\ge H_-$ then
$N_{-m^2}(H_+)\le N_{-m^2}(H_-)$. $N_{-m^2}(H_+)$ is an overestimate of the number of the bound states of $H_+$ {\it at} $-m^2$ for a fixed value of $\lambda$ 
but satisfactory for our purpose here.

By the Cwinkel-Lieb-Rozenblum bound in the form \cite{25a}
\begin{equation}\label{3.23}
N_{-m^2}(H_-)\le C\int d^3x\left[\left|\frac{e}{\lambda}\right||{\bf B}(x)|-m^2\right]_+^{3/2},
\end{equation}
where $[a]_+=\max(a,0)$ and $C=2\times 0.1156$. The factor 2 accounts for
the additional spin degrees of freedom in the present estimate.
Since $|\lambda_n|=O(1)$, we are confident that the degeneracy/multiplicity
associated with each $\lambda_n$ in (\ref{3.13}) does not exceed $c|e|^{3/2}\int d^3x|{\bf B}|^{3/2}$, where $c\ge 0.2312$ is another finite constant.  
This estimate has to be modified for values of $n>N$ beyond which
$\lambda_n$ assumes its asymptotic form as discussed below. Therefore,
for $n\le N$ we will estimate the sum in (\ref{3.13}) by factoring out
the common maximal degeneracy $c|e|^{3/2}\int d^3x|{\bf B}|^{3/2}$ and treat each $\lambda_n$ in the factored sum as having multiplicity equal to one.
Those $\lambda_n$, if any, that vanish as $e\to\infty$ give a
subdominant contribution to $\ln{\rm det}_2$ in (\ref{3.13}) since by inspection their contribution grows at most as $\lambda_n^2|e/\lambda_n|^{3/2}$.

We now turn to the large $e$ dependence of $\lambda_n$. From here on
we assume that $\psi_n$ is normalized to one. By C-invariance we may
assume $e>0$. Now consider the expectation value of (\ref{3.20}):
\begin{equation}\label{3.24}
<n|({\bf P}-e{\bf A})^2|n>-\frac{e}{\lambda_n}<n|{\bm\sigma}\cdot{\bf B}|n>=-m^2.\end{equation}
From (\ref{3.24}) if $<n|{\bm\sigma}\cdot{\bf B}|n>\,>0$ then $\lambda_n>0$ and {\it vice versa}. Therefore, we need only consider $\lambda_n>0$ and write
\begin{equation}\label{3.25}
\lambda_n=\left[\frac{<n|({\bf P}-e{\bf A})^2|n>}{e<n|{\bm\sigma}\cdot{\bf B}|n>}+\frac{m^2}{e<n|{\bm\sigma}\cdot{\bf B}|n>}\right]^{-1},
\end{equation}
where $<n|{\bm\sigma}\cdot{\bf B}|n>\ne 0$ as (\ref{3.24}) must be satisfied. The case $\lambda_n=0$ for some $n$ corresponding to $<n|{\bm\sigma}\cdot{\bf B}|n>=0$ can be ignored as $\lambda_n=0$ contributes nothing to $\ln{\rm det}_2$ in (\ref{3.13}). An easy estimate gives
\begin{equation}\label{3.26}
|(\psi_n,{\bm\sigma}\cdot{\bf B}\psi_n)|\le(\psi_n,|{\bf B}|\psi_n)\le\max_x|{\bf B}(x)|.
\end{equation}
Because ${\bf B}\in L^2$ and is assumed infinitely differentiable then
$\max\limits_x|{\bf B}|$ is finite. Hence, $<n|{\bm\sigma}\cdot{\bf B}|n>$ is a bounded function of $e$ and $n$.

Now consider the ratio $R_n=<n|({\bf P}-e{\bf A})^2|n>/e<n$ $|{\bm\sigma}\cdot{\bf B}|n>$ in (\ref{3.25}). The case $R_n{{\phantom{slsht}}\atop{\overrightarrow{\scriptstyle e\gg 1}}}0$ is ruled out since this implies $\lambda_n\to\infty$. The case $R_n{{\phantom{slsht}}\atop{\overrightarrow{\scriptstyle e\gg 1}}}\infty$ implies $\lambda_n\to 0$, which gives a subdominant contribution
to (\ref{3.13}) as discussed above. The final possibility is $1\le R_n<\infty$
for $e\to\infty$. The case $R_n\to 1$ for $e\to\infty$ happens if
$<n|({\bf P}-e{\bf A})^2|n>\sim e<n|{\bm\sigma}\cdot{\bf B}|n>$. Since    $\psi_n\in L^2$, $<n|({\bf P}-e{\bf A})^2-e{\bm\sigma}\cdot{\bf B}|n>=0$ implies ${\bm\sigma}\cdot({\bf P}-e{\bf A})\psi_n=0$. Now
this may happen for the ${\bf B}$-fields considered so far. But if we
exclude zero-mode supporting ${\bf B}$ fields \cite{25b} from our analysis it
cannot. By so doing we can exclude the case $\lambda_n=1-\delta_n(e)$,
$\delta_n(\infty)=0$. We will see below why this is necessary.

We proceed to estimate the strong coupling limit of $\ln{\rm det}_2$
in (\ref{3.13}). First, consider the sum for $n\le N$. We need only consider
$0<|\lambda_n|<1$ for all $e$, including $e=\infty$ as concluded above. Hence,
on factoring out the common maximal multiplicity of the $\lambda_n$ we get
\begin{equation}\label{3.27}
\lim_{e\gg 1}\left|\sum_{n=1}^N\ln(1-\lambda_n^2)\right|\le c_1e^{3/2}\int d^3x|{\bf B}|^{3/2},
\end{equation}
where $c_1$ is a constant and noting again that the eigenvalues
$\lambda_n\to 0$ are subdominant.

Since $\lambda_n\to 0$ for $n\to\infty$ and $1/2\le|\ln(1-\lambda_n^2)/\lambda_n^2|\le 3/2$ for $\lambda_n^2<1/2$ the absolute convergence of the series in 
(\ref{3.13}) requires $\sum\limits_{n=1}^{\infty}\lambda_n^2<\infty$.
Consider this sum for $n>N$ and indicate the degeneracy
factors $\mu_n$ explicitly:
\begin{equation}\label{3.28}
S\equiv\sum_{n>N}^{\infty}\mu_n(e)\lambda_n^2(e).
\end{equation}
We estimated from (\ref{3.23}) that $\mu_n\le c|e/\lambda_n|^{3/2}\int d^3x|{\bf B}|^{3/2}$. So
\begin{equation}\label{3.29}
S\le ce^{3/2}\int d^3x|{\bf B}|^{3/2}\sum_{{n>N}\atop{\rm no\ degeneracy}}^{\infty}|\lambda_n|^{1/2}<\infty.
\end{equation}
This implies that for $n>N$
\begin{equation}\label{3.30}
|\lambda_n(e)|=\frac{C_n(e)}{n^{2+\epsilon}},
\end{equation}
where $\epsilon>0$ and $C_n$ is a bounded function of $n$ and $e$ with
$\lim\limits_{e\to\infty}C_n(e)<\infty$. Otherwise $|\lambda_n|<1$ for any $n$ cannot be satisfied. Accordingly, the series in (\ref{3.29}) is uniformly convergent in $e$ by the Weierstrass M-test and so
\begin{equation}\label{3.31}
\lim_{e\to\infty}\left|\sum_{n>N}^{\infty}\ln(1-\lambda_n^2)\right|/e^{3/2}\le c_2\int d^3x|{\bf B}|^{3/2},
\end{equation}
where $c_2$ is a constant. From (\ref{3.13}), (\ref{3.27}) and (\ref{3.31}) we conclude
\begin{equation}\label{3.32}
\lim_{e\to\infty}\left|\ln{\rm det}_2(1-e\Delta_A^{1/2}{\bm\sigma}\cdot{\bf B}\Delta_A^{1/2})\right|/e^{3/2}\le c_3\int d^3x|{\bf B}|^{3/2},
\end{equation}
where $c_3$   is another constant.

As a check on (\ref{3.32}) refer to (\ref{3.5}). For ${\bf B}\in L^{3/2}({\mathbbm R}^2)$ we found \cite{9}
\begin{equation}\label{3.33}
\ln{\rm det}_{{\rm QED}_3}\ge -\frac{Ze^{3/2}}{6\pi}\int d^2x|B(x)|^{3/2},
\end{equation}
for $B(x)\ge 0$ or $B(x)\le 0$, $x\in{\mathbbm R}^2$. $Z$ is the dimension of the remaining space box. We know that $\ln{\rm det}_2\le 0$ and $\ln{\rm det}_{S{\rm QED}_3}\ge 0$ in (\ref{3.5}). Specializing (\ref{3.32}) to these ${\bf B}$ fields it is seen that
the strong coupling growth of $\ln{\rm det}_2$ is consistent with (\ref{3.33}).

Finally, if zero mode supporting ${\bf B}$ fields were allowed we
would have obtained $\ln{\rm det}_2{{\phantom{slsht}}\atop{{=}\atop{e\gg 1}}}O(e^{3/2}\ln\rho(e))$, $\rho(e){{\phantom{slsht}}\atop{{\to}\atop{e\gg 1}}}0$, since
when $|\lambda_n|{{\phantom{slsht}}\atop{\widetilde{\scriptscriptstyle e\gg 1}}}1-\delta_n(e)$, $\delta_n(\infty)=0$, the logarithm in (\ref{3.13}) gives an additional factor $\ln\delta_n$. As will be seen below the limit (\ref{1.4}) requires $\lim\limits_{e\to\infty}\ln{\rm det}_2/e^{3/2}={\rm finite}$ (or zero).

\vskip0,5cm

\subsection{Strong coupling limit of (\ref{3.18})}

It remains to estimate the large coupling limit of the last
term in (\ref{3.18}),
\begin{equation}\label{3.38}
\begin{array}{ccc}
\displaystyle I\equiv\frac{T}{\pi}\int\limits_{e{\mathscr B}}^{\infty}\frac{dM^2}{\sqrt{M^2-m^2}}\left(\ln{\rm det}_{{\rm QED}_3}+\frac{e^2||{\bf B}||^2}{24\pi\sqrt{M^2}}\right) \\
\displaystyle=\frac{T}{\pi}\int\limits_{e{\mathscr B}}^{\infty}\frac{dM^2}{\sqrt{M^2-m^2}}\left(-\frac{e^2}{4\pi}\int\frac{d^3k}{(2\pi)^2}|\hat{\bf B}(k)|^2\right. \\
\displaystyle\times\hskip-0,15cm\int\limits_0^1\hskip-0,15cm dz\frac{z(1-z)}{[z(1-z)k^2+M^2]^{\frac12}}\hskip-0,05cm+\hskip-0,05cm\frac{e^2||{\bf B}||^2}{24\pi\sqrt{M^2}}\hskip-0,05cm+\hskip-0,05cm\ln{\rm det}_4(1\hskip-0,05cm-\hskip-0,05cm eS\hskip-0,09cm\not\hskip-0,13cm A)\hskip-0,1cm\biggr), 
\end{array}
\end{equation}
\vskip0,2cm
\noindent where we substituted (\ref{2.6}) for $\ln{\rm det}_{{\rm QED}_3}$. Calculation 
of the first two terms in (\ref{3.38}) is straightforward. The last term has already been estimated in Sec. II and is given by the second term in (\ref{2.15}).
Hence,
\begin{equation}\label{3.39}
I=O\left(\frac{e^2T\int({\bf B}\cdot{\bf B})^2}{{\mathscr B}^2}\right)+O\left(\frac{eT\int{\bf B}\cdot\nabla^2{\bf B}}{{\mathscr B}}\right).
\end{equation}
Taking into account (\ref{3.32}) and (\ref{3.39}) we obtain from (\ref{3.18})
\begin{equation}\label{3.40}
\lim_{e\to\infty}\frac{\ln{\rm det}_{ren}}{e^2\ln e}\ge\frac{||{\bf B}||^2T}{24\pi^2}.
\end{equation}
Equations (\ref{2.16}) and (\ref{3.40}) therefore establish (\ref{1.4}).

\vskip0,5cm

\section{SUMMARY}

The two assumptions underlying (\ref{1.4}) are first that the continuum limit of the lattice diamagnetic inequality coincides with (\ref{1.5}), and second that the heat kernel expansion of the Pauli operator in (\ref{2.14}) is an asymptotic series. These assumptions can and should be proven or falsified.

In addition, the result (\ref{1.4}) assumes that the vector potential
and magnetic field satisfy the following conditions:

${\bf B}\in L^2({\mathbbm R}^3)$ to define $\ln{\rm det}_{ren}$ in (\ref{2.5}) and to ensure that $\Delta_A^{1/2}{\bm\sigma}\cdot{\bf B}\Delta_A^{1/2}\in{\mathscr I}_2$ following Appendix B. In addition ${\bf B}\in L^{3/2}({\mathbbm R}^3)$ in order that the degeneracy estimate in (\ref{3.23}) is defined. To ensure that the bound in (\ref{3.32}) holds, zero mode supporting ${\bf B}$ fields are excluded. Also {\bf B} must be infinitely
differentiable $(C^{\infty})$ to ensure that the expansion coefficients in
(\ref{2.14}) are finite.

If {\bf A} is assumed to be in the Coulomb gauge then by (\ref{2.3}) 
${\bf A}\in L^6 ({\mathbbm R}^3)$. If ${\bf B}\in L^{3/2}({\mathbbm R}^3)$ then ${\bf A}\in L^3({\mathbbm R}^3)$ by the Sobolev-Talenti-Aubin inequality \cite{11}. In order to define ${\rm det}_{{\rm QED}_3}$
it is necessary to assume ${\bf A}\in L^r ({\mathbbm R}^3),\ r > 3$,
following the discussion under (\ref{2.6}). If ${\bf A}\in L^3({\mathbbm R}^3)$ and $L^6({\mathbbm R}^3)$, then  ${\bf A}\in L^r({\mathbbm R}^3),\ 3<r<6$ also. This follows from 
H${\ddot{\rm o}}$lder's inequality \cite{25}

\setcounter{equation}{0}
\renewcommand{\theequation}{4.\arabic{equation}}

\begin{equation}\label{4.1}
||fg||_r\le||f||_p||g||_q,
\end{equation}
with $p^{-1}+q^{-1}=r^{-1},\ p,q,r\ge 1$. Since ${\bf B}={\bm\nabla}\times{\bf A}$ and
${\bf B}\in C^{\infty}$ then ${\bf A}\in C^{\infty}$.

We note that the sample functions $A_\mu(x)$ supporting the Gaussian
measure in (\ref{1.2}) with probability one are not $C^{\infty}$. It is generally accepted that they belong to ${\mathscr S}'({\mathbbm R}^4)$, the space of tempered distributions. Therefore, we point out here that the $C^{\infty}$ functions we introduced can be related to $A_\mu\in{\mathscr S}'({\mathbbm R}^4)$ by the convoluted field $A_\mu^\Lambda(x)=\int d^4y\,f_\Lambda(x-y)A_\mu(y)\in C^{\infty}$, provided $f_\Lambda\in{\mathscr S}({\mathbbm R}^4)$, the functions
of rapid decrease. Then the Fourier transform of the
covariance $\int d\mu(A)A_\mu^\Lambda(x)A_\nu^\Lambda(y)$ derived from (\ref{1.2}) is $\hat D_{\mu\nu}(k)|\hat f_\Lambda(k)|^2$, where $\hat f_\Lambda\in C^{\infty}$. Since ${\rm QED}_4$ must be ultraviolet regulated before
renormalizing, $\hat f_\Lambda$ can serve as the regulator by choosing, for
example, $\hat f_\Lambda=1,\ k^2\le\Lambda^2$ and $\hat f_\Lambda=0,\ k^2\ge 2\Lambda^2$. So the need to regulate can serve as a natural way to introduce $C^{\infty}$ background fields $A_\mu^\Lambda$ into ${\rm det}_{ren}$ -- but not the rest of $d\mu$ in (\ref{1.1}) -- and into
whatever else one is calculating. This procedure is a generalization
of that used in the two-dimensional Yukawa model \cite{26}.

Finally, the obvious generalization of (\ref{1.4}) for an admissible
class of fields on ${\mathbbm R}^4$ is
\begin{equation}\label{4.3}
\lim_{e\to\infty}\frac{\ln{\rm det}_{ren}}{e^2\ln e}=\frac{1}{48\pi^2}\int d^4x\,F_{\mu\nu}^2(x).
\end{equation}
There is no chiral anomaly term since $F_{\mu\nu}$ falls off faster than
$1/|x|^2$ and $\int d^4x\,\tilde F_{\mu\nu}F_{\mu\nu}=\int d^4x\,\partial_\alpha(\epsilon_{\alpha\beta\mu\nu}A_\beta F_{\mu\nu})=0$., where
$\tilde F_{\mu\nu}=\frac12\epsilon_{\mu\nu\alpha\beta}F_{\alpha\beta}$. Equation (\ref{4.3}) remains to be verified.

If (\ref{1.4}) and (\ref{4.3}) do indeed indicate instability then they are
yet another reason why QED should not be considered in isolation.

\vskip0,5cm

\centerline{\bf APPENDIX A}

\vskip0,5cm

The operator indentity on which (\ref{3.1}) is based is obtained as
follows. Let \cite{27}
$$
F_t=e^{-t(X+Y)}e^{tX}.
$$
Then
$$
\frac{dF_t}{dt}=-e^{-t(X+Y)}Ye^{tX}.
$$
Integrating gives

\setcounter{equation}{0}
\renewcommand{\theequation}{A\arabic{equation}}

\begin{equation}\label{A1}
e^{-t(X+Y)}-e^{-tX}=-\int\limits_0^tds\,e^{-(t-s)(X+Y)}Ye^{-sX},
\end{equation}
known as Duhamel's formula. Iterating once gives the required identity:
\begin{eqnarray}\label{A2}
e^{-t(X+Y)}-e^{-tX}=-\int\limits_0^tds\,e^{-(t-s)X}Ye^{-sX} \nonumber \\
+\int\limits_0^t\hskip-0,1cm ds_1\hskip-0,1cm\int\limits_0^{t-s_1}\hskip-0,1cm ds_2e^{-(t-s_1-s_2)(X+Y)}Ye^{-s_2X}Ye^{-s_1X}.
\end{eqnarray}

\vskip0,5cm

\centerline{\bf APPENDIX B}

\vskip0,5cm

Here we show that the operator $K=\Delta_A^{1/2}{\bm\sigma}\cdot{\bf B}\Delta_A^{1/2}\in{\mathscr I}_2$ and
hence that $K$ is Hilbert-Schmidt. This follows \cite{13,14,25} if and only if $K$ is a bounded operator on $L^2({\mathbbm R}^3,d^3x;{\mathbbm C}^2)$ having a representation of the form

\setcounter{equation}{0}
\renewcommand{\theequation}{B\arabic{equation}}

\begin{equation}\label{B1}
(Kf)(x)=\int {\cal K}(x,y)f(y)d^3y,\ f\in L^2,
\end{equation}
where
\begin{equation}\label{B2}
{\cal K}(x,y)=<x|\Delta_A^{1/2}{\bm\sigma}\cdot{\bf B}\Delta_A^{1/2}|y>,
\end{equation}
and where ${\cal K}\in L^2(\mathbbm R^3\times\mathbbm R^3;d^3x\times d^3y)$. Moreover,
\begin{equation}\label{B3}
||K||_2^2=\int|{\cal K}(x,y)|^2d^3xd^3y.
\end{equation}
If it can be shown that ${\cal K}\in L^2$ then it trivially follows that $K$ maps $L^2$ into itself. So consider
\begin{eqnarray}\label{B4}
||{\cal K}||_{L^2}=2\sum_i\int d^3xd^3yB_i(x)\Delta_A(x,y)B_i(y)\Delta_A(y,x) \nonumber \\
\le 2\sum_i\int d^3xd^3y|B_i(x)||\Delta_A(x,y)||B_i(y)||\Delta_A(y,x)|.\quad
\end{eqnarray}
A form of Kato's inequality \cite{4,20,28} asserts that the interacting scalar
propagator is bounded by the free propagator
\begin{equation}\label{B5}
|\Delta_A(x,y)|\le\Delta(x-y),
\end{equation}
where $\Delta(x)=(4\pi|x|)^{-1}e^{-m|x|}$ in three dimensions. Then
\begin{equation}\label{B6}
||{\cal K}||_{L^2}\le\frac{1}{8\pi^2}\sum_i\hskip-0,1cm\int\hskip-0,1cm d^3xd^3y|B_i(x)|\frac{1}{|x-y|^2}e^{-2m|x-y|}|B_i(y)|.
\end{equation}
By Young's inequality in the form \cite{22}
\begin{equation}\label{B7}
\left|\int d^3xd^3yf(x)g(x-y)h(y)\right|\le||f||_p||g||_q||h||_r,
\end{equation}
where $p^{-1}+q^{-1}+r^{-1}=2,\ p,q,r\ge 1$, and $||f||_p=(\int d^3x|f(x)|^p)^{1/p}$ etc., obtain from (\ref{B6}) with $p=r=2,\ q=1$
\begin{equation}\label{B8}
||{\cal K}||_{L^2}\le\frac{1}{8\pi^2}\sum_i||B_i||^2\hskip-0,1cm\int\hskip-0,1cm d^3x\,e^{-2m|x|}/x^2=||{\bf B}||^2(4\pi m)^{-1}.
\end{equation}
\vskip0,2cm
\noindent Therefore, by the theorem that began this appendix
$\Delta_A^{1/2}{\bm\sigma}\cdot{\bf B}\Delta_A^{1/2}\in{\mathscr I}_2$
when $m\ne 0$ and ${\bf B}\in L^2$. We mention that this can be proved even when $m=0$ provided ${\bf B}\in L^{3/2}$.

\end{document}